\documentclass[journal]{IEEEtran}
\usepackage{cite}
\usepackage{amsmath,amssymb,amsfonts}
\usepackage{algorithm}
\usepackage{algorithmic}
\usepackage{graphicx}
\usepackage{textcomp}
\usepackage{xcolor}
\usepackage{multirow}
\usepackage{soul}
\usepackage{subcaption}
\usepackage[inkscapelatex=false]{svg}
\usepackage{url}
\ifCLASSINFOpdf
\else
\fi
\begin{document}
%
\title{A Diffusion-based Generative Machine Learning Paradigm for Dynamic Contingency Screening}
%
%
%

\author{Quan~Tran,~\IEEEmembership{Student Member,~IEEE,}
        Suresh~S.~Muknahallipatna,~\IEEEmembership{Senior Member,~IEEE,} 
        Dongliang~Duan,~\IEEEmembership{Senior Member,~IEEE,}
        Nga~Nguyen,~\IEEEmembership{Senior Member,~IEEE.}
\thanks{Quan Tran, Suresh S. Muknahallipatna, Dongliang Duan, and Nga Nguyen are with the Department of Electrical Engineering and Computer Science, University of Wyoming, Laramie, WY, 82071 USA (e-mail: qtran2@uwyo.edu or hongquandnpc@gmail.com; sureshm@uwyo.edu; dduan@uwyo.edu; and nga.nguyen@uwyo.edu).}
}
\maketitle

\begin{abstract}
Dynamic contingency screening is a challenging task in dynamic security assessment, when traditional numerical approaches are computationally intensive and often not able to repeatedly solve full AC power flow for all possible contingencies in real time, especially for large-scale power grids. Moreover, the severity caused by a contingency is not identical for all operating points, which does not necessitate solving all possible contingencies computationally inefficient and time-consuming. This paper introduces a novel, diffusion-based generative machine learning paradigm that transforms contingency analysis from conventional scenario selection to a proactive, likely-unsupervised scenario generation. The margin to the steady-state voltage stability limit determines the ranking of contingencies corresponding to each operating point. By leveraging physical information from each operating point, the proposed approach anticipates the contingencies most likely to be critical, without relying on static assumptions or exhaustive simulations. This data-prompted generative approach enables the identification of high-risk scenarios under varying load and generator conditions, providing dynamic security assessment in real time. The correctness, effectiveness, and scalability of the methodology are demonstrated through methodological derivations and comprehensive experiments on multiple IEEE benchmark systems, including IEEE-6, IEEE-14, IEEE-30, and IEEE-118 \footnote{Code and data are publicly published at: \url{https://github.com/hongquandnpc/Diffusion4DynamicContingencyScreening.git}}, highlighting its potential to incorporate contingency screening in complex, evolving smart grids. 
\end{abstract}


%
\IEEEpeerreviewmaketitle

\section{Introduction}
\label{intro}
Contingency screening is an indispensable part of the dynamic security assessment, with the ultimate goal of finding contingencies that would lead to instability and easily damageable regions in power systems \cite{mansourDynamicSecurityContingency1997}. It is also a significant concern when ensuring static security power systems in the context of diversified perturbations that have emerged increasingly, such as the fast-paced integration of renewable energy resources and electric vehicles. This circumstance poses more challenges for system operators when operating power systems in such an unanticipated, complicated environment, but the requirements of stable condition operations are always on top. A potential instability of systems probably originates only from minor facts, such as a drastic voltage drop at a system's bus, out of numerous buses of large-scale power systems. Hence, preventive and corrective control actions that are determined from contingency screening are essentially well-prepared. However, it is time-consuming and resource-wasting to investigate all possible contingencies. Therefore, it is necessary to innovate a technique to enhance the screening of contingencies effectively and accurately.


\IEEEpubidadjcol

A significant domain of contingency screening is an aspect of the related-voltage security ranking of contingencies. Along with the performance index relating to power flows, another index for voltage-reactive power performance was defined in \cite{albuyeh1982reactive}, which was calculated from the first iteration of the Fast Decoupled Power Flow before applying the selection of credible contingencies. A local solution method in \cite{lauby1983contingency} that was a simplified version of the concentric relaxation method \cite{zaborszky1980fast}, executed in a screening process to determine a voltage increment larger than a predetermined value, would be solved by full AC power flow. Likewise, two different voltage performance indices were proposed in \cite{nara1985line} to yield the ranking of contingencies before using a linearized decoupled model to solve them for finding violations. Similarly, these methods employed various methods of contingency screening for sorting the ranks to reduce considerable numbers of cases in order to alleviate the computational burden of solving the full AC power flow. Nonetheless, the computation of full AC power flow was still required though the number of cases was partly reduced by algorithms of contingency selection.


Recently, machine learning algorithms have been state-of-the-art techniques that have been dominant in multiple realms of technology and disciplines. Nevertheless, in terms of the field of contingency screening, the term ``artificial intelligence system" was employed earlier. Particularly, an expert system for screening contingency in power systems was proposed in \cite{sobajic1988artificial}. It was constructed on predefined rules to perform as a filter to load flow to inspect highly risky contingencies and endangered regions. Its achievement was significant as to decrease the number of contingencies with the adequate effectiveness of screening. However, superficial knowledge, inflexibility, and the closed system are prominently inherent disadvantages of an expert system \cite{aleksandrovich2021hybrid}. To cover the drawbacks of the expert system, the work in \cite{mansour1997dynamic} deployed neural networks to screen and rank dynamic security contingencies. It utilized a simple three-layer neural network with an input of given power system information and an output of energy margin and swing angle. Likewise, a similar structure of an artificial neural network was employed in \cite{schafer2018contingency} with an input of given information of a power system, and an output of power flow information. The architecture of a multilayer perceptron was used to yield the predicted results of contingencies. Though the achieved results were adequate, the proposed model \cite{schafer2018contingency} was only employed in a fixed topology, and it did not cover all power system parameters as well as large perturbations such as dropping off a line or a generator.

The proposed paradigm leverages the diffusion theory and image-processing generative models to generate the most critical contingencies through learning the probability distribution hidden in data patterns. In essence, it is a generative model that produces new novel data based on the learned distribution from training data. Its principle includes three processes: the forward process, the backward process, and the sampling process. The validation is implemented on 
three case studies of IEEE-6, 
IEEE-14, IEEE-30, and IEEE-118 to illustrate the proposed paradigm's performance and efficacy. Basically, this paper has the following contributions as follows:
\begin{itemize}
    \item It proposes a novel mindset for contingency screening by generating instead of selecting. This paper develops a new approach to the application of machine-learning algorithms that leverages the diffusion generative model to generate the worst contingencies. 
    \item A comprehensive index to quantify the severity of contingencies based on the risk of voltage collapse is proposed based on the continuation power flow method. 
    \item The proposed paradigm introduces the modification of the original diffusion algorithms to make the diffusion theory applicable to a specific problem of power systems. 
    This work reduces the computational burden of the contingency screening process and shorten significantly the computational time for tasks that require an instant response.
\end{itemize}

This paper is organized as follows. Section \ref{method} presents the methodology of the proposed paradigm. Section \ref{diffusion} describes the implementation and algorithms of the proposed paradigm for contingency screening. Section \ref{results} illustrates the simulation results and discussions. Finally, Section \ref{conclusion} concludes the paper.

\section{The methodology}
\label{method}
\subsection{The continuation power flow}
\label{cpf}

The continuation power flow algorithm (CPF) finds a continuum of power flow solutions from a base case, aiming to a target case with higher scheduled power \cite{ajjarapu1992continuation}. The algorithm makes the anticipation of the increased level of injection power and then corrects that prediction by using its results to obtain the appropriate increase in power. Therefore, the CPF is often known as a predictor-corrector method and classified as a general class of path-following methods \cite{bryant1990equilibrium,rheinboldt1986numerical}. It may trace the power system's steady-state behavior to the variation of loads and generators. Furthermore, it can overcome the difficulties of conventional power flow algorithms, which are not able to determine a solution in the vicinity of saddle-node bifurcation points by reformulating the power flow equations into a set of augmented ones in order to maintain the well-conditioned Jacobian matrix at various loading or generating levels \cite{kundur2007power}.


Generally speaking, by augmenting a continuation parameter $\lambda$ to the power flow equations, the CPF is briefly described as follows \cite{zimmerman2016matpower}:
\begin{equation}
  (x^j, \lambda^j) \xrightarrow{\text{Predictor}} (\hat{x}^{j+1}, \hat{\lambda}^{j+1}) \xrightarrow{\text{Corrector}} (x^{j+1}, \lambda^{j+1})
\end{equation}

In particular, the power flow equations are incorporated with a load parameter $\lambda$ as follows \cite{ajjarapu1992continuation}:

\[0 \le \lambda \le \lambda_{critical} \]

where the base case is respective to $\lambda = 0$ and $\lambda = \lambda_{critical}$ corresponds to the critical injection power.

\begin{align}
    0 &= P_{Gi} - P_{Li} - P_{Ti} \\
    \label{eq:2}
    0 &= Q_{Gi} - Q_{Li} - Q_{Ti} \\
    \label{eq:3}
    P_{Ti} &= \sum\limits_{j=1}^{n} V_{i}V{j}y_{ij}\cos(\delta_{i} - \delta_{j} - \nu_{ij}) \\
    Q_{Ti} &= \sum\limits_{j=1}^{n} V_{i}V{j}y_{ij}\sin(\delta_{i} - \delta_{j} - \nu_{ij})
\end{align}
    
where:
\begin{itemize}
    \item[-] $i, j$ is the notation of bus i, 
    \item[-] $L, G, T$ is the notation of bus load, generator, and injection,
    \item[-] $V\angle \delta_{i}, V\angle \delta_{j}$ is the voltage at bus i and bus j,
    \item[-] $y\angle\nu_{ij}$ is the $(i,j)^{th}$ element of $Y_{bus}$.
\end{itemize}

The critical state in which the limit of maximum transferable power amount is reached is determined from a nose curve. This curve is a plot of increasing/decreasing loading levels versus the voltage variations. To quantify the steady state loading limit due to the load variations, a power transfer schedule is specified to provide a power base that is appropriate for scaling of $\lambda$, the load parameter.

The equations \eqref{eq:2} and \eqref{eq:3} are rewritten as follows:
\begin{equation}
  g(x) = 
    \begin{bmatrix}
    P(x) - P_{T} \\
    Q(x) - Q_{T}
    \end{bmatrix}
    = \begin{bmatrix}
    0 \\
    0
    \end{bmatrix}
\label{eq:6}
\end{equation}

where: $P(x)$ and $Q(x)$ represent the power variation of loads and generators with respect to $P_{T}$ and $Q_{T}$ are the active/reactive power injections. 

With an augmented parameter $\lambda$ into the power flow equations, the equation \eqref{eq:6} can be restructured as:
\begin{equation}
  f(x, \lambda) = g(x) - \lambda b = 0
  \label{eq:7}
\end{equation}
With: $x \equiv (\Theta, V_m)$ and
\begin{equation*}
  b = 
  \begin{bmatrix}
  P_{T}^{target} - P_{T}^{base} \\
  Q_{T}^{target} - Q_{T}^{base} 
  \end{bmatrix}
\end{equation*}

where: $b$ is the vector of scheduled power transfer. Within the paper's scope, $b$ is determined by the deficit of a base case and a target case that is defined by fixed values throughout the process of conducting experiments.

By parameterizing the values of $(x, \lambda)$ along the nose curve by the methods in \cite{chiang1995cpflow,li2008nonlinear}, the tangent vector $z^j = \left[ dx \, d\lambda \right]^T_j$ at step $j$ is obtained from the augmented equations as follows \cite{zimmerman2016matpower}:
\begin{equation}
  \begin{bmatrix}
  f_x & f_\lambda \\
  p_{x}^{j-1} & p_{\lambda}^{j-1}
  \end{bmatrix}
  z^j =
  \begin{bmatrix}
  0 \\
  1
  \end{bmatrix}
\end{equation}
where: $p^j(x,\lambda)$ is one of three parameterization schemes, i.e., Natural parameterization, Arc length parameterization, and Pseudo arc length parameterization \cite{zimmerman2016matpower}.

The predictor stage is completed after updating the current values of $(x, \lambda)$ by the tangent vector $\overline{z}^j$, which is normalized by its L2 norm:
\begin{equation}
\overline{z}^j = \frac{z^j}{\|z^j\|_2}
\end{equation}
\begin{equation}
  \begin{bmatrix}
  x^{j+1} \\
  \lambda^{j+1}
  \end{bmatrix}
  =
  \begin{bmatrix}
  x^j \\
  \lambda^j
  \end{bmatrix}
  + \alpha \overline{z}^j
\end{equation}

where: $\alpha$ is the step size.

The corrector stage is executed by using Newton's method to find the next solutions based on modifying the previous approximation ones $\left(\hat{x}_{i+1}, \hat{\lambda}_{i+1}\right)$ yielded by the predictor stage. The next solutions $\left(x_{i+1}, \lambda_{i+1}\right)$ are obtained by solving the parameterized power flow equations of \eqref{eq:7} added the parameterization constraint \cite{zimmerman2016matpower}:
\begin{equation}
\begin{bmatrix}
f(x, \lambda) \\
p^j(x, \lambda)
\end{bmatrix}
= 0
\end{equation}

The sign of the tangent differential term of $\lambda$ (i.e., $d\lambda$) implies the impact of varying the load parameter on the voltage profile. In particular, the step whose its sign of $d\lambda$ is positive on the upper portion of the $V-P$ curve is zero at the saddle-bifurcation point, and is negative on the lower portion. Hence, the alternating sign of $d\lambda$  indicates whether the system reaches the equilibrium point or not \cite{kundur2007power}.

\subsection{Mathematical technique to detect voltage collapse}
\label{indicator}

\begin{figure*}[!t]
  \centering
  \centerline{\includegraphics[width=1.9\columnwidth]{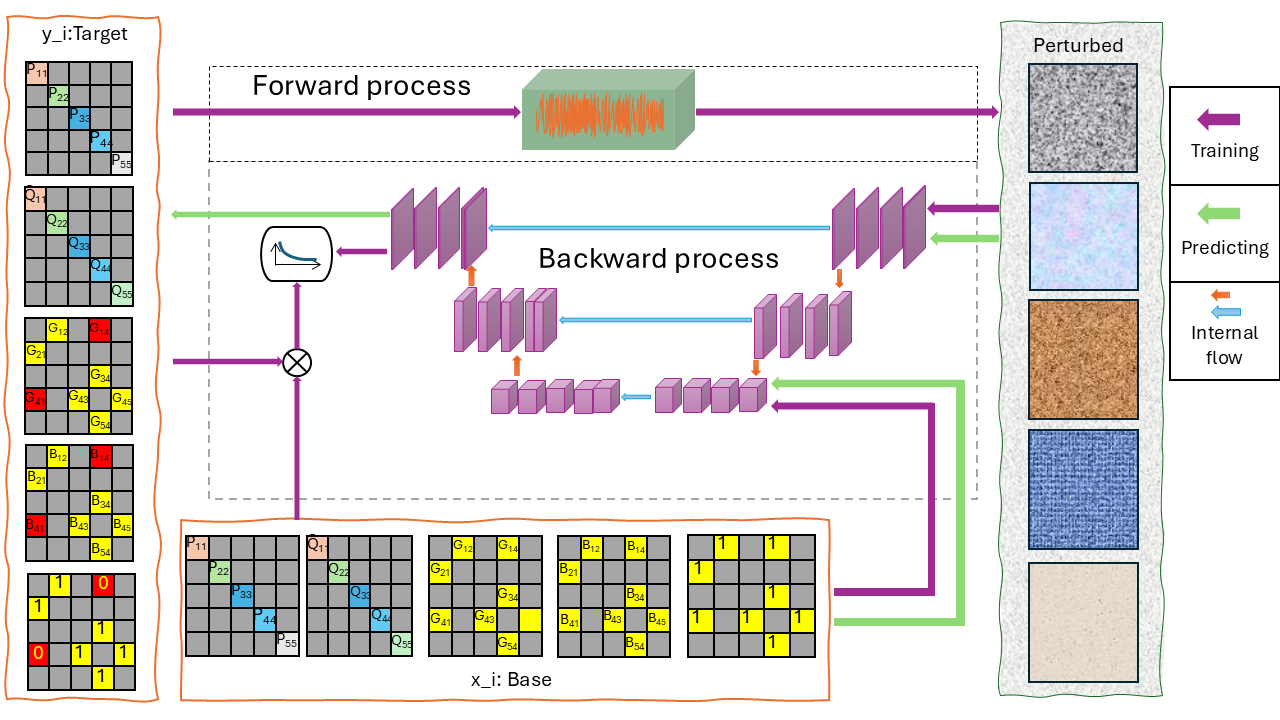}}
  \caption{The illustration of the proposed model DDPM-CS: The forward process adds noise to the critical operating point $y\_i$, which originates from a system snapshot $x\_i$, until the perturbed state is obtained. The reverse process utilizes a U-NET neural network with an encoded prompt of a system snapshot $x\_i$ added to the bottleneck to estimate the added noise. Finally, the sampling procedure generates new data based on the encoded prompt.}
  \label{fig:diffusion}
\end{figure*}

Voltage stability is one of the most concerning problems that is frequently discussed. In the heavily stressed electric systems, transmission lines or generator units are operated close to their limits. This situation may result from a common contingency of the increasing load demand, along with a line outage in practice. Several contingency events of that kind may be extreme enough to put the system at risk of voltage collapse. To this end, voltage collapse is mostly admitted as a major motivation for multiple systems' collapse events \cite{IEEEspecialpublication,talor1993modelling,mansour1991voltage,10935083,powersystemvoltagestability}. Therefore, tracing equilibrium points of the systems is of primary interest for opting an appropriate method to determine voltage collapse points \cite{van2007voltage,ajjarapu1998bibliography}.

Mathematically, small or large disturbances may be visualized as nondeterministic changes, and the power systems are considered dynamically as being involved in those disturbances. The power system dynamics are represented by a set of vector-form differential equations \cite{vu1995voltage}:

\begin{equation}
  \dot{\mathbf{x}} = f(\mathbf{x}, \lambda)
  \label{eq:12}
\end{equation}

where: 
\begin{itemize}
  \item[-] $\mathbf{x}$ is the state vector of the system, i.e., angles/magnitudes of voltage; 
  \item[-] $\lambda$ is the parameter that indicates the causes of disturbances, such as taking load demand or the working status of elements, into account.
\end{itemize}

Small disturbance is a common phenomenon in the system, typically when load demand varies in a tight range, not far from the base demand. In other words, the pre-disturbance and post-disturbance equilibria are close to each other. The equation, which demonstrates a change due to a `small' fluctuation of load demand, is derived from equation \eqref{eq:12}. It is a linearized form of the system's dynamics revolving around a pre-disturbance equilibrium point $\mathbf{x}_0$ \cite{vu1995voltage,hossain2014robust}: 
\begin{equation}
  \Delta \dot{\mathbf{x}} = A(\mathbf{x}_0, \lambda) \Delta \mathbf{x}
\end{equation}

where: $A(\mathbf{x}_0, \lambda)$ is the state matrix at the equilibrium point $\mathbf{x}_0$ and $\lambda$.
Based on the time-dependent characteristic of a mode with respect to the state matrix's eigenvalues, the system's stability is determined. A system is sufficiently stable for small disturbances if all of one's eigenvalues have negative real parts, which indicate a decaying mode \cite{kundur2007power}.

What about large disturbances? Large disturbances is all of things remained in the set of things called disturbances. In other words, they are significant events - known as contingencies - that may lead a system towards a significant change in the system's state, i.e., transmission line outages or the loss of generator units.  In this case, the variation range of $\lambda$ is substantially large and unpredictable. Theoretically, these cases is likely to be dictated by numerical methods as small disturbance cases. Nevertheless, the computation burden required to be executed is not tractable for large-scale power systems in the real world. This is also a major motivation of this paper to shift to a novel approach of contingency screening. It will detect the worst scenarios of power systems by leveraging the perturbation diffusion technique beside other methods utilizing Lyapunov theory in \cite{chang1995direct}. It would be discussed in detail in the next section. 

The numerical methods, which were mentioned earlier, center around the accomplishment of solutions from the power-flow balance equations. The most common method is the Newton-Raphson iterative technique, which derives multiple variants, such as Decouple Power Flow or Fast Decouple Power Flow. Its variants are applied to effectively alleviate the computational burden for rapidly determining critical contingencies \cite{wood2013power,davis2010multiple,vykuka2015sensitivity}. However, the serious challenge of these methods is the matter of Jacobian singularity at the bifurcation point or the steady state voltage stability limit. Therefore, a locally parameterized continuation technique was proposed in \cite{ajjarapu1992continuation}, known as the continuation power flow (CPF). The fundamental background of the CPF is introduced in Subsection \ref{cpf}. Due to its merit, it is chosen as a reliable numerical tool to determine the direct margin calculation to a point of voltage collapse in this paper.




\subsection{Diffusion in Machine Learning}

In the context of machine learning, diffusion models are a class of latent variable generative models to achieve novel data \cite{sohl2015deep}. A general diffusion model performs a stochastic process that includes adding noise to data in a forward process and denoising noise in a reverse process, given a set of training datasets. At the inference stage, the trained model uses random noise as its input to generate new data.

The structure of a diffusion model is generally represented by the forward process, the reverse process, and the sampling procedure \cite{chang2023design}. The generic workflow is performed by a timestep-indexed multi-step chain executed by the forward process and the reverse process. The sampling procedure is conducted by the reverse chain to generate new data from the noise. As introduced in \cite{ho2020denoising}, the forward and reverse process are implemented together to train a denoising network by gradually adding and removing noise. The sampling procedure is executed by utilizing the trained model through the forward and reverse process to generate a novel a sample of data $x_0^* \sim p_\theta(x_0) \approx p(x_0)$ \cite{song2019generative}.

The forward process transforms an input from an original entity, such as an image, to a fully noisy entity that is totally different from the original one. During a forward chain involving multiple consecutive timesteps $t$, a Gaussian noise $\epsilon_t$ whose magnitude is controlled by a hyperparameter $\beta_t$, is added to $x_{t-1}$. Specifically, the forward process is represented by the following equation \cite{sohl2015deep,ho2020denoising}:
\begin{align}
  q(x_T | x_0) & := q(x_1 | x_0) \cdots q(x_t | x_{t-1}) \cdots q(x_T | x_{T-1}) \nonumber \\
               & := \prod_{t=1}^{T} q(x_t | x_{t-1})
  \label{eq:14}
\end{align}
where: $x_0 \sim q(x_0)$ is the data distribution; $q(x_t | x_{t-1})$ is the transition by adding a noise following a Gaussian distribution \cite{ho2020denoising}; $t$ is the timestep index; $T$ is the total number of time steps. Likewise, variational autoencoders (VAEs) \cite{kingma2013auto}, the forward process performs a perturbation on the input data to generate noisy data whose distribution is also a Gaussian distribution \cite{prince2023understanding}.

The reverse process performs the denoising of the noisy data generated by the forward process. To perform the opposite transition, a neural network is trained to learn the noise that is added before the forward process. The reverse process executes a backward chain $p_{\theta} (x_{t-1} | x_t)$ to denoise the noisy data by gradually removing the noise that is inferred from the neural network. The reverse process is described by the following equation \cite{sohl2015deep,ho2020denoising}:
\begin{align}
  p_\theta(x_{0:T}) & := p(x_T) p_\theta(x_{T-1} | x_T) \cdots p_\theta(x_{t-1} | x_t) \cdots p_\theta(x_0 | x_1) \nonumber \\
          & := p(x_T) \prod_{t=1}^{T} p_\theta(x_{t-1} | x_t)
  \label{eq:15}
\end{align}
where: the joint distribution $p_{\theta} (x_{0:T})$ is the reverse process. This process acts as a backward Markov chain, with the starting state $x_T$ being a random Gaussian noise. The ending state $x_0$ results from the denoising process by utilizing the learned noise from the neural network parameterized by $\theta$. 

The sampling procedure essentially leverages the reverse process to generate novel data $\hat{x}^*_0 \sim p_{\theta}$ with the identical sequence. The optimized neural network represented by $\theta$ acts as a noise predictor that may produce noise that is most similar to the noise added during the forward process. As a result, the sampling procedure is analogous to the reverse process:
\begin{align}
  p_{\theta} (\hat{x}^*_{0:T}) & := p(x_T) p_{\theta} (x_{T-1} | x_T) \cdots p_{\theta} (x_0 | x_1) \nonumber \\
          & := p(x_T) \prod_{t=1}^{T} p_{\theta} (x_{t-1} | x_t)
  \label{eq:16}
\end{align}


\subsection{The fundamental mathematical background}

The mathematical derivation is presented in detail in \cite{strumke2023lecture}; thus, this section will briefly introduce the results used to derive a novel foundation formula for the loss function, which underpins the main algorithm in this work.

To be continued, the resulting generation is definitely undoing the diffusion forward process. An appropriate neural network is leveraged to execute this reverse process by learning on how Gaussian noise is added by the forward process. The neural network model is trained to learn the distribution of the training data by maximizing $p_{\theta}(\mathbf{x}_{0})$, the likelihood of the data points $\mathbf{x}_0$, where the likelihood is produced by the model parameterized by $\theta$. 
\begin{equation}
  p_\theta(\mathbf{x}_0) = \int p_\theta(\mathbf{x}_{0:T}) d\mathbf{x}_{1:T}
    \label{eq:17}
\end{equation}
where: $p_\theta(\mathbf{x}_{0:T})$ is defined by Equation \ref{eq:15}, i.e., the joint distribution of the reverse process. It can be said that the likelihood is a marginal distribution over all time steps from $t=T$ to $t=1$ from the expression in Equation \ref{eq:17}.

The maximization of the likelihood is equivalent to the minimization of the negative log-likelihood. In other words, the loss function, which is used to train the model, can be written as a negative log-likelihood:
\begin{equation}
  \mathcal{L}(\theta) = -\log p_\theta(\mathbf{x}_0)
  \label{eq:18}
\end{equation}

The logarithmic and integral operations in Equation \ref{eq:18} are intractable and calculated analytically \cite{strumke2023lecture}. A derivation by utilizing the Evidence Lower Bound, which is a lower bound on $\log p_{\theta}{\left( \textbf{x}_0 \right)}$ is presented:
\begin{align}
\log &p_\theta(\mathbf{x}_0) \geq \mathbb{E}_{q(\mathbf{x}_1|\mathbf{x}_0)} \left[\log p_\theta(\mathbf{x}_0|\mathbf{x}_1)\right] \nonumber \\
& - \mathbb{E}_{q(\mathbf{x}_{T-1}|\mathbf{x}_0)} \left[\mathcal{D}_{\text{KL}} \left(q(\mathbf{x}_T|\mathbf{x}_{T-1})||p_\theta(\mathbf{x}_T)\right)\right] \nonumber \\
& - \sum_{t=1}^{T} \mathbb{E}_{q(\mathbf{x}_{t-1,t+1}|\mathbf{x}_0)} \left[\mathcal{D}_{\text{KL}} \left(q(\mathbf{x}_t|\mathbf{x}_{t-1})||p_\theta(\mathbf{x}_t|\mathbf{x}_{t+1})\right)\right]
\label{eq:19}
\end{align}

Finally, the loss function is expressed as:

\begin{align}
    -\log &p_\theta(\mathbf{x}_0) \leq \mathcal{L}_{\text{vlb}} := \sum_{t=2}^{T} \mathcal{L}_{t-1} \nonumber \\
    &\quad = \sum_{t=2}^{T} \frac{1}{2\tilde{\beta}_t} \frac{\bar{\alpha}_{t-1} \cdot \beta_t^2}{(1 - \bar{\alpha}_t)^2} \cdot \mathbb{E}_{q(\mathbf{x}_t|\mathbf{x}_0)} \left[ \left\| \hat{\mathbf{x}}_\theta(\mathbf{x}_t, t) - \mathbf{x}_0 \right\|_2^2 \right]
    \label{eq:20}
\end{align}
where: $\beta_t$ is a variance whose noise is gradually added at time $t$ of the forward process; $\bar{\alpha}_t := \prod_{s=1}^{t} \alpha_s$ with $\alpha_s = 1 - \beta_s$, and $\hat{\mathbf{x}}_\theta(\mathbf{x}_t, t)$ is the denoised state with respect to the noisy state $\mathbf{x}_t$ and time $t$.

In practice, the model training only requires the optimization of the expectation term in Equation \ref{eq:20} and neglects the weighting term as conducted empirically by \cite{ho2020denoising}:
\begin{equation}
  \mathcal{L}_{t-1} \leftarrow  \mathbb{E}_{q(\mathbf{x}_t|\mathbf{x}_0)} \left[ \left\| \hat{\mathbf{x}}_\theta(\mathbf{x}_t, t) - \mathbf{x}_0 \right\|_2^2 \right]
  \label{eq:21}
\end{equation}

Equation \ref{eq:21} is reformulated in \cite{ho2020denoising} under the term of noise rather than an original data point $\mathbf{x}_0$, interpreted as learning the noise added from the forward process to denoise from the noisy object back to approximately an original data points:  
\begin{equation}
  \mathcal{L}_{t-1} \leftarrow  \mathbb{E}_{q(\mathbf{x}_t|\mathbf{x}_0)} \left[ \left\| \hat{\mathbf{\epsilon}}_\theta(\mathbf{x}_t, t) - \mathbf{\epsilon}_t \right\|_2^2 \right]
  \label{eq:22}
\end{equation}
where: $\mathbf{\epsilon}_t$ is the noise added at time $t$ of the forward process; $\hat{\mathbf{\epsilon}}_\theta(\mathbf{x}_t, t)$ is the predicted noise achieved from the reverse process or denoising process.

\subsection{Proposed loss function for contingency generation aware of distinctive base profiles}
\label{lossfunction}

As presented in Section \ref{method}, the load parameter $\lambda$ obtained from the continuation power flow is considered as a metric to determine the relative distance to the voltage instability. It can be explained further that the detrimental effect of each contingency is different for each base profile due to the various amplitudes of $\lambda$ as relative distances from different base case profiles to target case ones at which voltage instability appears or the system is on the edge of the boundary of instability on the way to be asymptotical to target cases.

From the definition of the vector of scheduled power transfer in Equation \ref{eq:7}, we redefine as following to a concept viewed on the training dataset's standpoint between a base case and a target case:

\textbf{Definition 1.} A base case and a target case are defined as follows:
\begin{itemize}
  \item A base case is a normal operational condition of a power system, which needs to figure out a contingency whose consequence may cause the worst effect to the system, such as voltage collapse.
  \item A target case is structured as the base case, a condition of a power system, at which power profiles at each bus are assumed to be a state of the system whose operations are ideally at most. In the scope of this work, all generators are arbitrarily allowed to operate at most $10\%$ of the nominal power generation output, meanwhile all bus loads are arbitrarily assigned at most $500\%$ of the power demand of the original case study's values. 
\end{itemize}

Noted that the loadscale of $500\%$ serves as an adjustable hyperparameter for the system to compute the load parameter $\lambda$ in a broad range of training datasets. In other words, it can be set arbitrarily as long as it suffices to have as many of the data points whose continuation power flow solutions converged from base cases as possible.

\textbf{Definition 2.} The element-wise subtraction between a target case and a base case is defined as the discrepancy tensor $\mathbf{\xi}$ between the two corresponding entries of those cases. 
\begin{itemize}
  \item $\mathbf{\xi}_0$ is a tensor of the true data point given by $\left( \mathbf{x}_0,\mathbf{x}^*_0 \right)$, where $\mathbf{x}_0$ and $\mathbf{x}^*_0$ are the profiles of a base case and a target case at time $t=0$, respectively.
  \begin{equation}
    \xi_0 := \mathbf{x}^*_0 - \mathbf{x}_0
    \label{eq:23}
  \end{equation}
  \item $\hat{\mathbf{\xi}}_\theta \left( \mathbf{x}_t,t \right)$ is a tensor of the predicted outcome given by $\left( \hat{\mathbf{x}}_\theta(\mathbf{x}_t, t), \mathbf{x}^*_0 \right)$, where $\hat{\mathbf{x}}_\theta(\mathbf{x}_t, t)$ is the denoised state obtained from the reverse process with respect to the noisy state $\mathbf{x}_t$ obtained from the forward process at time $t$.
  \begin{equation}
    \hat{\mathbf{\xi}}_\theta \left( \mathbf{x}_t,t \right) := \mathbf{x}^*_0 - \hat{\mathbf{x}}_\theta(\mathbf{x}_t, t)
    \label{eq:24}
  \end{equation}
\end{itemize}

To derive an appropriate loss function for the development of a novel diffusion-based generative model for contingency screening, we state the following theorem to provide the mathematical foundation:

\textbf{Theorem 1.} Let $\theta$ denote for a parameterized neural network model learning the reverse process. Without losing the correctness and completeness of the original likelihood maximization in Equation \ref{eq:17} for training a diffusion-based neural network model, 

The loss function of a diffusion-based neural network model for contingency screening is to minimize the L2 norm expectation in term of the loss between $\hat{\xi}_\theta \left( \mathbf{x}_t,t \right)$ and $\xi_0$. In a similar manner of Equation \ref{eq:21}, the novel loss function is determined as follows:
\begin{align}
  \mathcal{L}_{t-1} &\leftarrow \mathbb{E}_{q(\mathbf{x}_t|\mathbf{x}_0)} \left[ \left\| \mathbf{\xi_0} - \hat{\mathbf{\xi}}_\theta(\mathbf{x}_t, t) \right\|_2^2 \right]
  \label{eq:25}
\end{align}

\textbf{Proof.} Consider Equation \ref{eq:21}, where $\mathbf{x}_0$ denotes the true profile of a base case and $\hat{\mathbf{x}}_\theta(\mathbf{x}_t, t)$ represents the predicted profile of the denoised state. We manipulate the loss function in Equation \ref{eq:21} through an algebraic transformation that preserves the mathematical equivalence of the optimization objective without losing the initial correctness and completeness of the likelihood maximization in Equation \ref{eq:17}.

Specifically, we introduce the target case profile $\mathbf{x}^*_0$ by adding and subtracting it within the norm:

\begin{align}
  \mathcal{L}_{t-1} &\leftarrow  \mathbb{E}_{q(\mathbf{x}_t|\mathbf{x}_0)} \left[ \left\| \hat{\mathbf{x}}_\theta(\mathbf{x}_t, t) - \mathbf{x}_0 \right\|_2^2 \right] \nonumber \\
  &\leftarrow \mathbb{E}_{q(\mathbf{x}_t|\mathbf{x}_0)} \left[ \left\| \hat{\mathbf{x}}_\theta(\mathbf{x}_t, t) - \mathbf{x}^*_0 + \mathbf{x}^*_0 - \mathbf{x}_0 \right\|_2^2 \right]
  \label{eq:26}
\end{align}

This transformation enables us to reframe the optimization in terms of the loss between base and target cases. Following \textbf{Definition 2}, we define:
\begin{align}
  \mathbf{\xi} &:= \mathbf{x}^*_0 - \mathbf{x}_0 \nonumber \\
  \mathbf{\hat{\xi}}_\theta(\mathbf{x}_t, t) &:= \mathbf{x}^*_0 - \hat{\mathbf{x}}_\theta(\mathbf{x}_t, t)
  \nonumber
\end{align}

Substituting these definitions into Equation \ref{eq:26}, we obtain the reformulated loss function as stated in \textbf{Theorem 1}:
\begin{align}
  \mathcal{L}_{t-1} &\leftarrow \mathbb{E}_{q(\mathbf{x}_t|\mathbf{x}_0)} \left[ \left\| \mathbf{\xi_0} - \hat{\mathbf{\xi}}_\theta(\mathbf{x}_t, t) \right\|_2^2 \right]
  \label{eq:27}
\end{align}

Equation \ref{eq:27} provides the theoretical foundation for extending the original diffusion framework of \cite{ho2020denoising} to a physics-informed, diffusion-based generative learning paradigm tailored for contingency screening. Accordingly, from this point forward, the subsequent adaptation of the Denoising Diffusion Probabilistic Model (DDPM) \cite{ho2020denoising} to the context of contingency screening is both methodologically justified and scientifically well-grounded.

Nevertheless, the nature of contingency generation differs fundamentally from that of image generation, as it must satisfy and be validated against the strict physics-based constraints. Therefore, developing a novel generative model for this task requires not only an appropriate loss formulation but also substantial modifications and the explicit incorporation of physical information into the training process, in addition to the correct loss function. The proposed algorithms that address these requirements are presented in detail in Section \ref{ddpm_implementation}.

\section{Diffusion Model for Contingency Screening}
\label{diffusion}

\subsection{$N-1$ criterion in Contingency Screening}

Contingency screening is one of three primary activities to execute the online security assessment \cite{morante2006pervasive}. Additionally, power system security analysis is challenging to deal with large-scale electrical power networks composed of numerous interconnected equipment, i.e., generators, and power transformers, which are connected by transmission lines. The transmission networks are objects that are prone to most risk vulnerability by unanticipated impacts of the environments in which they are operated. Therefore, under the unforeseen contingencies caused by transmission lines, the system's stability is put at more risk of being compromised. Especially, in the case of a crucial tie-line, the system is likely to reach to the risk of collapse due to the sudden change of network structure, which may lead to the loss of power balance \cite{di2004distributed,aloisio1997distributed}.

The $N-1$ criterion is a common approach to dealing with power system security analysis \cite{morante2006pervasive}. It is defined as a line outage or a generator/transformer failure occured in a power system, but the system is still able to maintain its stability \cite{schafer2018contingency}. Because of being more easily influenced by external factors of the transmission lines than the other equipment, a single line outage is commonly considered as the $N-1$ criterion. Therefore, the $N-1$ criterion is utilized to validate a novel method of contingency screening by a diffusion-based generative machine learning paradigm in this paper. The utilization of this criterion would be helpful in evaluating the performance of the proposed model, as it is one of the most concerning issues in practical operational activities. 

\subsection{The proposed model's principle}
\label{model_principle}

The restructuring of power grid information is of an image-like shape is proven to be beneficial for applying machine learning algorithms in power systems. Load demand and grid topology information are organized as 2D matrices; the power profile matrix is a diagonal matrix whose elements represent the power profile, while the connection matrix's elements correspond to bus connections in the topology.  With this image-like structure, the paradigm of the convolutional neural network family (CNN) is leveraged to approximate alternate current optimal power flow with acceptable accuracy and high-speed computation \cite{tran2024learning,tran2024advanced,tran2025ac}. The information organization is demonstrated in detail in Section \ref{data_acquisition}. 

That point is also a motivation for this work regarding searching for the most detrimental contingencies by machine-learning-based paradigms. As discussed in Section \ref{intro}, the ultimate goal of contingency screening is to pick several worst contingencies with the least effort. It is highly significant for assembling schemes of preventive and corrective instant actions to maintain the system's stability. However, it is a challenging task to determine the worst scenarios of power systems by numerical analytics due to the large-scale, dynamic, and complex nature of power systems. 

\begin{figure}[!h]
  \centering
  \centerline{\includegraphics[width=0.9\columnwidth]{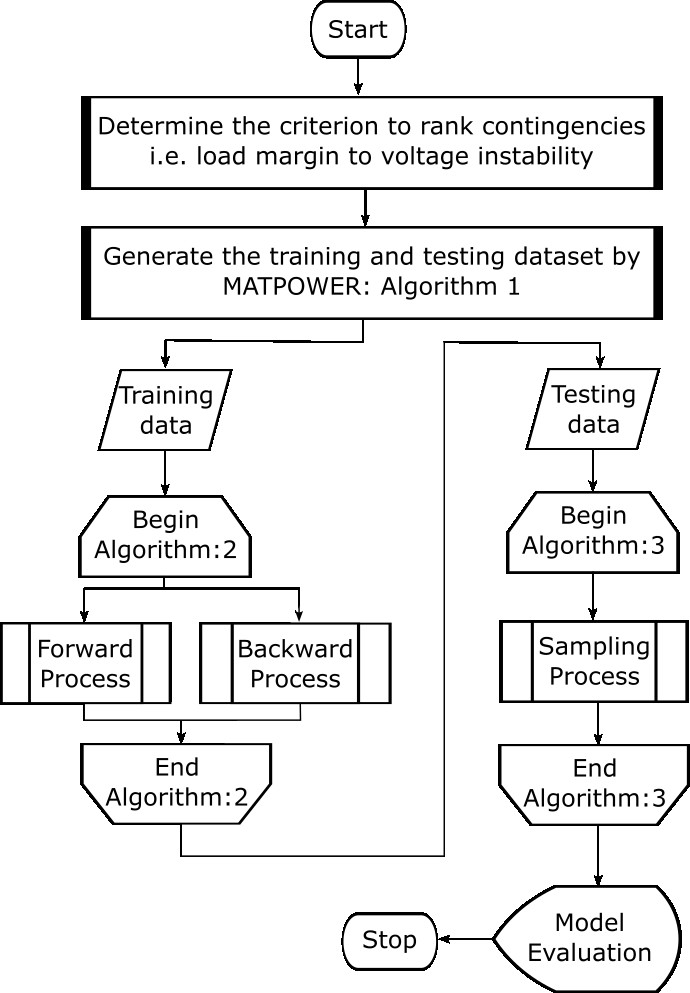}}
  \caption{The flowchart of the DDPM-CS's implementation.}
  \label{fig:flowchart}
\end{figure}

Inspired by Denoising Diffusion Probabilistic Models - DDPM \cite{ho2020denoising}, the proposed model is a novel approach for screening contingencies. Specifically, a highly risky system state that is prone to instability is restructured as an original image. The original one will be fully noisy by a forward process, then denoised by a reverse process trained to learn the noise added by the forward process. The trained model is utilized to generate a new image or a new likely-worst state from random noise. In short, the principle of the proposed model is illustrated in Fig.~\ref{fig:flowchart}.

\subsection{Data generation algorithm}

The data generation algorithm \ref{alg:1} is designed to create a dataset of power system states that are prone to instability for the case studies IEEE-6, 
IEEE-14, IEEE-30, and IEEE-118. The algorithm is executed by perturbing the load demand and the incidence matrix of the power system from a base state. The perturbation is performed by multiplying a random Gaussian noise that is scaled to $[0.5, 1.5]$ to the grounded load demand. A random $N-1$ contingency of line outages is selected following the uniform distribution, which is represented by a connection matrix $\textbf{C} = [c_{ij}]$. The geometrical configurations of lines in the case studies are known as the topologies, demonstrated as non-directional graphs in which two buses are linked by a branch, are said to be connected. Specifically, the connection matrix $\textbf{C}$ is defined as follows:

\begin{align}
  c_{ij} = 
  \begin{cases} 
      0 & \text{if bus } i \text{ is not connected to bus } j \\
      1 & \text{if bus } i \text{ is connected to bus} j \\
  \end{cases}
\end{align}
 
In the scope of this paper, $(\max\_\lambda)_i$ is the maximum value of the load parameter $\lambda$ to which a power transfer schedule can be scaled without causing voltage instability. It is computed by the continuation power flow method as presented in Section \ref{cpf}. In a quantitative manner, it is prescribed as the direct margin calculation or the indicator of a voltage instability, which is discussed in Section \ref{indicator}. If the power flow analysis converges, the critical load demand at a saddle bifurcation point and the connection matrix with respect to the randomly-chosen $N-1$ contingency are added to the dataset. The algorithm repeats this process for a specified number of samples $N$ and selects the top 10\% samples with the lowest $\max\_\lambda$ such that the proposed model is physically-informed from the worst contingencies. To this end, the proposed model may learn essential features hidden in the data patterns and grid structures, and since then, it may generate the most detrimental scenarios of power systems from any base state of power systems.

\begin{algorithm}[H]
\caption{Data Generation}
\begin{algorithmic}[1]
\STATE Initialize the case studies.
\STATE Define the number of samples $N$
\FOR{$i = 1, \ldots, N$}
  \STATE Randomly perturb load demand and connection matrix from base state $x_i$:
    \begin{itemize}
      \item[-] Load demand $\sim \mathcal{N} \left (0, 1 \right )$ scaled to $\left [0.5, 1.5 \right ]$
      \item[-] Connection matrix $\triangleq$ \{ Select a $N-1$ contingency $\sim \text{Uniform}\{lines\}$ \}
    \end{itemize}
  \STATE Run continuation power flow analysis
  \IF{converged}
    \STATE Extract $\left (\max\_\lambda \right )_i$ and $y_i$, including load demand at the critical point and the new connection matrix w.r.t. the $N-1$ contingency.
  \ELSE
    \STATE Discard the sample
  \ENDIF
\ENDFOR
\STATE Sort the dataset by $\max\_\lambda$ in ascending order
\STATE Select the top 10\% samples with the lowest $\max\_\lambda$
\STATE \textbf{return} dataset $\{(x_1,y_1), \ldots, (x_n,y_n)\}$ where: $n=10\%\text{N}$
\end{algorithmic}
\label{alg:1}
\end{algorithm}

\subsection{The DDPM-CS's implementation}
\label{ddpm_implementation}

The proposed model is motivated by the Denoising Diffusion Probabilistic Models (DDPM) \cite{ho2020denoising}, which is a class of latent variable generative models. Additionally, it is physically informed by utilizing power grids' instinct features to generate a novel sample of input data, i.e., power system states that are prone to instability. Thus, the proposed model is considered a DDPM-based model, named the Denoising Diffusion Probabilistic Model for Contingency Screening (DDPM-CS). Nevertheless, it is not completely identical to the original DDPM due to their significant difference in their own purposes. Instead of generating a new image from random noise, the DDPM-CS is trained to generate a novel data sample, whose predicted power system states are straightforwardly prone to voltage instability. In other words, the outcome of the DDPM-CS is the worst scenarios of power systems without the need to conduct a large-scale numerical analysis of all probable contingencies in the power systems.

The DDPM-CS's framework is basically similar to the original DDPM, which encompasses the forward process, the reverse process, and the sampling procedure. Nonetheless, the training algorithm \ref{alg:2} is constructed on the proposed loss function, which is presented in Section \ref{lossfunction}, distinct from that of the original DDPM to tailor the proposed model to converge under strict physics-based constraints of power systems. In particular, the target case $\mathbf{x}^*_0$ is added by a Gaussian noise during the timestep $1\ldots T$ by a given variance schedule $\beta_1, \ldots, \beta_T$ to generate the noisy data $\mathbf{x}_t$ at each timestep $t$. The Gaussian noise, which is controlled by the given schedule, is gradually added during the timestep until the original target case $\mathbf{x}^*_0$ is fully perturbed to a noisy data $\mathbf{x}_T \sim \mathcal{N} \left ( \textbf{0}, \textbf{I} \right )$ \cite{ho2020denoising}. As a result, the whole process of perturbing the target case $\mathbf{x}^*_0$ to a fully noisy data $\mathbf{x}_T$ is defined as the forward process, which is represented by Algorithm \ref{alg:2}.

After the perturbation of the input data $x_0$ to a fully noisy form $x_T$ by the forward process, the reverse process is defined by leveraging a neural network architecture to denoise the noisy data $x_T$ to a novel sample $\mathbf{x}_{t-1|t=1}$ whose distribution is closely approximated to the true one of $\mathbf{x}^*_0$. The distribution of the novel-generated samples is a distribution $f_\theta(x)$ that is asymptotically approximated to the original distribution $f(x)$ by the training on the neural network. For the sake of efficiency, the U-Net architecture \cite{ronneberger2015u} is adapted to the DDPM-CS for the denoising function by the reverse process. 

Theoretically, the U-Net architecture is a family of convolutional neural networks that is engineered as a common structure of an encoder followed by a decoder to learn the noise added by the forward process. The U-Net architecture is a U-shaped encoder-decoder that is symmetric and composed of a series of convolutional layers and skip connections. There is an information bottleneck in the middle of the U-Net architecture to reduce the dimensionality of the feature maps and engage the network to learn features (noise) from the prompt effectively. To serve the role of denoising noise, after being trained to learn the discrepancy at each time $t$ of the timestep with respect to a base profile, the U-Net model infers the discrepancy $\mathbf{\xi}_\theta \left( \mathbf{x}_t,t \right)$ between the noised version of the target case $\mathbf{x}_t$ and the base case $\mathbf{x}_0$. 

The denoising process is repeated over the time step from $t=T$ to $t=1$ to generate a novel sample $\mathbf{\hat{x}}^*_0 \sim p_\theta(\mathbf{\hat{x}}^*_0)$, where $p_\theta(\mathbf{\hat{x}}^*_0)$ is approximate to the true distribution $p(\mathbf{x}_0)$, i.e. the distribution of the worst scenarios with respect to different operating base cases in the training dataset. Nevertheless, in this proposed model, the denoising process wears a completely different manner. In particular, the sampling process leverages a novel approach based on the new designed loss function to obtain the denoised version of target cases at each time $t$, which is derived directly from a discrepancy $\hat{\mathbf{\xi}}_\theta \left (\mathbf{x}_t, t \right )$ as illustrated in step 4 of Algorithm \ref{alg:3}. Consequently, that of target cases at $t-1$ is a result of the process of adding noise for time $t-1$ from that denoised version at time $t$, which is exactly what the forward process makes the target cases noisy. The sampling procedure is described in Algorithm \ref{alg:3}.

\begin{algorithm}[H]
\caption{Training}
\begin{algorithmic}[1]
\REPEAT
  \STATE \textbf{Forward Process}
  \STATE $ \mathbf{x}^*_0 \sim f \left (\mathbf{x_0} \right )$
  \STATE $\mathbf{\xi} = \text{loss} \left (\mathbf{x}^*_0, \mathbf{x_0} \right )$
  \STATE $t \sim \text{Uniform} \left (\{1, \ldots, T\} \right )$
  \STATE $\mathbf{x}_t \sim f \left (\mathbf{x}_t; \sqrt{\bar{\alpha}_t} \mathbf{x}^*_0, \left (1 - \bar{\alpha}_t \right )\mathbf{I} \right )$

  \STATE \textbf{Reverse Process}
  \STATE $\mathbf{\xi}_{\theta} \sim f_{\theta} \left (\mathbf{x}_t \right )$
  \STATE Take gradient descent step on $\nabla_\theta \left \| \mathbf{\xi_0} - \hat{\mathbf{\xi}}_\theta \left (\mathbf{x}_t, t \right ) \right \|^2$
\UNTIL{converged}
\end{algorithmic}
\label{alg:2}
\end{algorithm}

The DDPM-CS algorithms differ from the original DDPM in both the loss function and the sampling process, although they are built on the original DDPM. The resulting innovation drives the proposed model to be a novel diffusion-based generative model for a specific task in power systems, which is to generate the most detrimental scenarios of power systems from any base state of power systems. This capability is critical because power system operations are inherently dynamic, and the system's stability is likely to be compromised by the unanticipated contingencies. However, exhaustively screening all possible contingencies for each operating point is computationally prohibitive, especially when accounting for rapid variations in load demand. The proposed approach's results after conducting experiments on the case studies, which are presented in Section \ref{results}, provide strong evidence of its efficiency and the empirical validity.

\begin{algorithm}[H]
\caption{Sampling}
\begin{algorithmic}[1]
\STATE $\mathbf{x}_T \sim \mathcal{N} \left (\textbf{0}, \textbf{I} \right )$
\FOR{$t = T, \ldots, 1$}
  \STATE $\mathbf{z} \sim \mathcal{N} \left (\textbf{0}, \textbf{I} \right )$ if $t > 1$ else $\mathbf{z} = \textbf{0}$
  \STATE $\mathbf{x}_t = \mathbf{x}_0 + \hat{\mathbf{\xi}}_\theta \left (\mathbf{x}_t, t \right )$
  \STATE $\mathbf{x}_{t-1} = \sqrt{\bar{\alpha}_{t-1}} \cdot \mathbf{x}_t + \sqrt{1 - \bar{\alpha}_{t-1}} \cdot \textbf{z}$ if ${t > 1}$ else $\mathbf{x}_t$
\ENDFOR
\STATE \textbf{return} $p_\theta(\mathbf{\hat{x}}^*_0) \approx p(\mathbf{x}^*_0)$
\end{algorithmic}
\label{alg:3}
\end{algorithm}

\section{Experiment Results} 
\label{results}
\subsection{Data Acquisition}
\label{data_acquisition}

The training dataset generated by the data generation algorithm \ref{alg:1} is executed by MATPOWER \cite{zimmerman2016matpower}. The target values of active/reactive power of load demand and generators $(P^{target}_T, Q^{target}_T)$ in the equation \eqref{eq:7} are determined by the fixed scaled values of the base case $(P^{base}_T, Q^{base}_T)$ for all case studies, i.e. IEEE-6, 
IEEE-14, IEEE-30, and IEEE-118. The case studies' dataset is available in MATPOWER's library.

Generally speaking, the training dataset is utilized to train the Denoising Diffusion Probabilistic Models for Contingency Screening (DDPM-CS) by the training algorithm \ref{alg:2}. The model is evaluated by the testing dataset according to the criterion of the contingency ranking. Specifically, a novel generated contingency is validated by MATPOWER to determine its value of $\max\_\lambda$. As a result, the DDPM-CS performance is demonstrated by the ranking of the generated contingencies over all possible ones of the same sample.

\subsection{Results and Discussion}

A threshold value is introduced as the median of the total number of contingencies of a case study. As discussed in Subsection \ref{indicator}, the rank of a contingency per sample in the testing dataset is determined by the value of $\max\_\lambda$ computed by the continuation power flow method (MATPOWER). The threshold value is utilized to assess comprehensively the DDPM-CS's performance across all the case studies. The worst contingencies of which are of more interest are ranked from the lowest to the threshold value. The remaining contingencies are ranked higher than the threshold value. Due to this classification, the DDPM-CS's performance is evaluated by the score of the number of contingencies that are ranked under the threshold.  The DDPM-CS architecture is fine-tuned through the trial-and-error process to achieve appropriate hyperparameters in proportion to the case study's complexity.

\begin{figure*}[!t]
  \centering
  \begin{subfigure}[b]{0.85\columnwidth}
    \centering
    \includegraphics[width=\textwidth]{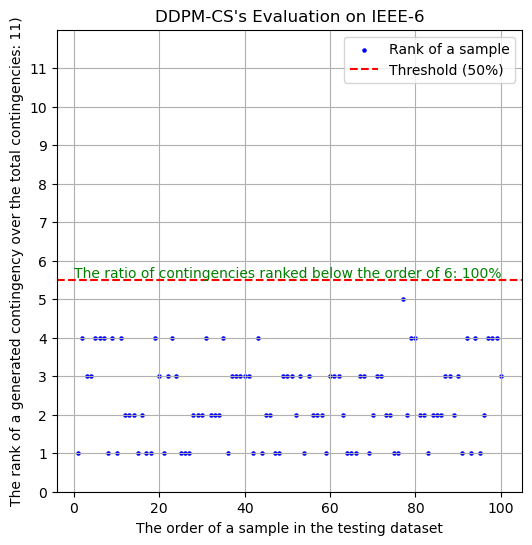}
    \label{fig:evaluation_ieee6}
  \end{subfigure}
  \hfill
  \begin{subfigure}[b]{0.85\columnwidth}
    \centering
    \includegraphics[width=\textwidth]{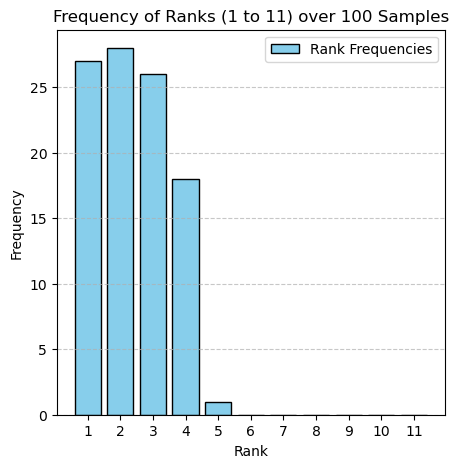}
    \label{fig:freq_ieee6}
  \end{subfigure}
  \caption{Evaluation and frequency analysis for IEEE-6 case study across 100 samples.}
  \label{fig:ieee6_analysis}
\end{figure*}


\begin{figure*}[!h]
  \centering
  \begin{subfigure}[b]{0.85\columnwidth}
    \centering
    \includegraphics[width=\textwidth]{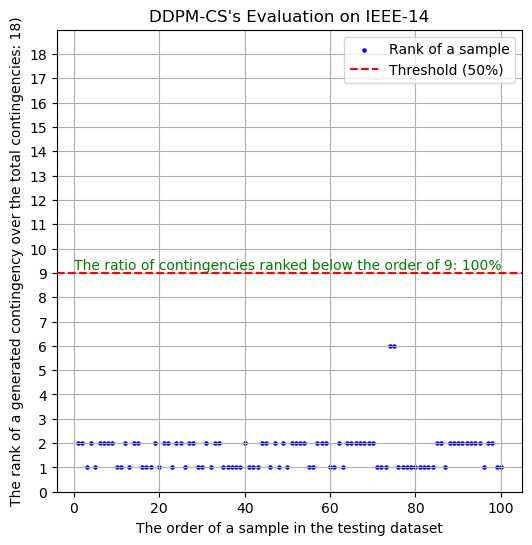}
    \label{fig:evaluation_ieee14}
  \end{subfigure}
  \hfill
  \begin{subfigure}[b]{0.85\columnwidth}
    \centering
    \includegraphics[width=\textwidth]{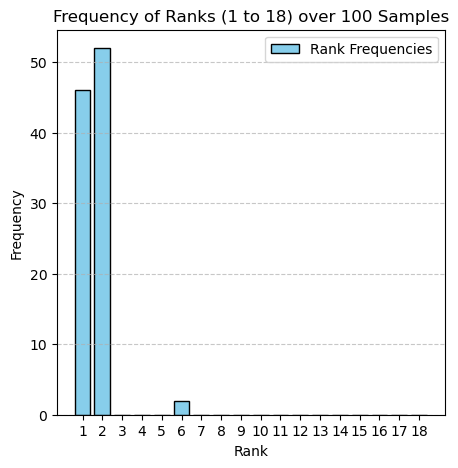}
    \label{fig:freq_ieee14}
  \end{subfigure}
  \caption{Evaluation and frequency analysis for IEEE-14 case study across 100 samples.}
  \label{fig:ieee14_analysis}
\end{figure*}

\begin{figure*}[!h]
  \centering
  \begin{subfigure}[b]{0.85\columnwidth}
    \centering
    \includegraphics[width=\textwidth]{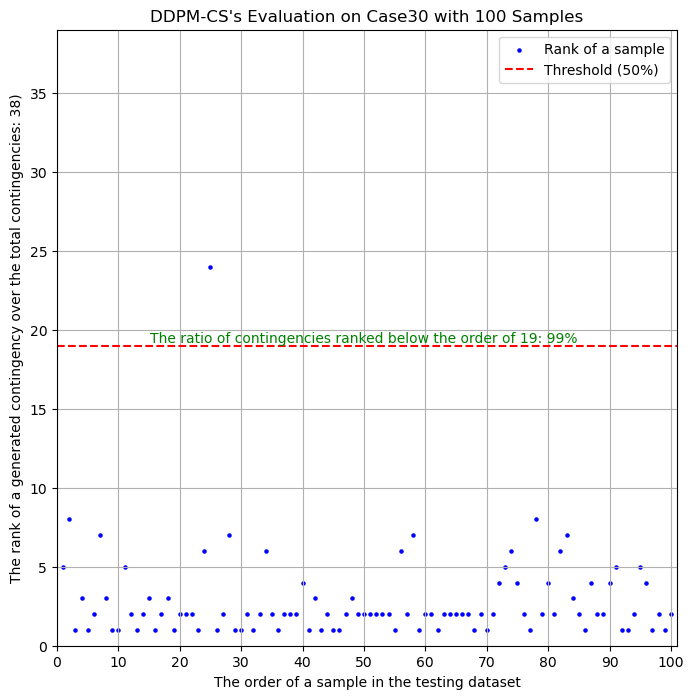}
    \label{fig:evaluation_ieee30}
  \end{subfigure}
  \hfill
  \begin{subfigure}[b]{0.85\columnwidth}
    \centering
    \includegraphics[width=\textwidth]{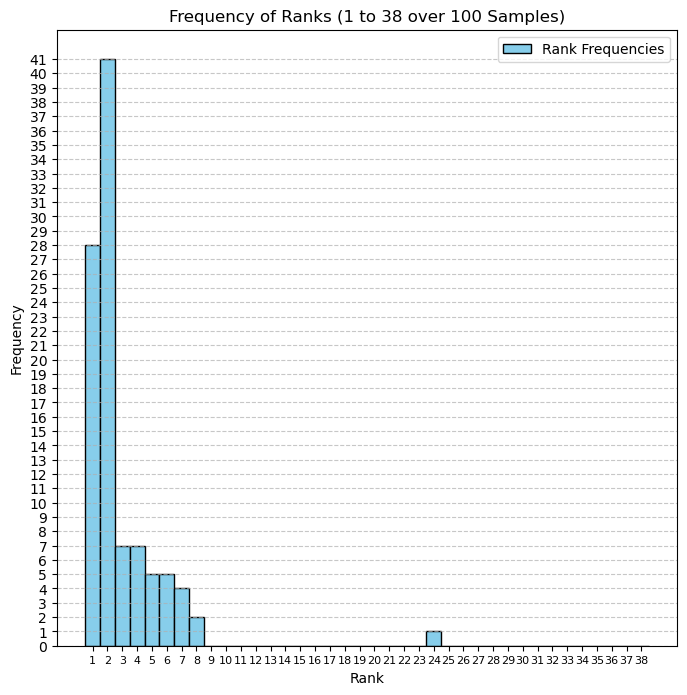}
    \label{fig:freq_ieee30}
  \end{subfigure}
  \caption{Evaluation and frequency analysis for IEEE-30 case study across 100 samples.}
  \label{fig:ieee30_analysis}
\end{figure*}

\begin{figure*}[!h]
  \centering
  \begin{subfigure}[b]{0.85\columnwidth}
    \centering
    \includegraphics[width=\textwidth]{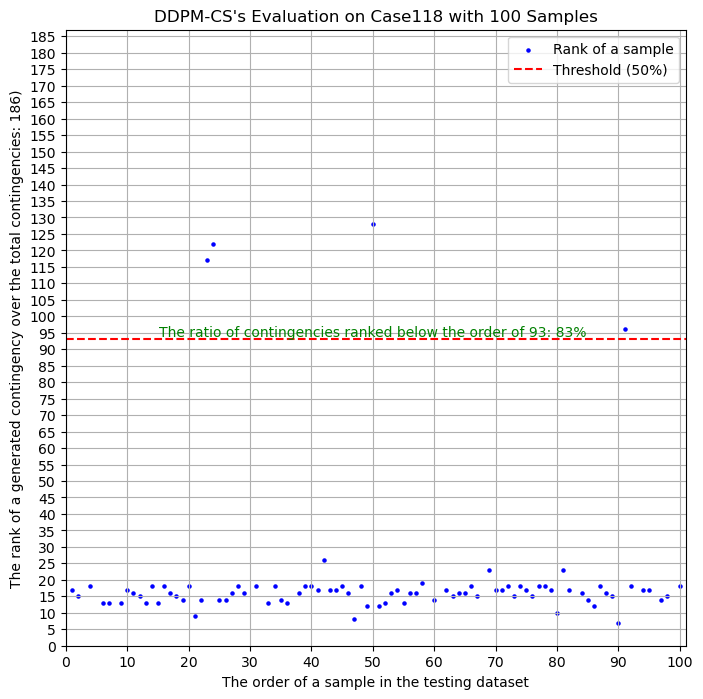}
    \label{fig:evaluation_ieee30}
  \end{subfigure}
  \hfill
  \begin{subfigure}[b]{0.85\columnwidth}
    \centering
    \includegraphics[width=\textwidth]{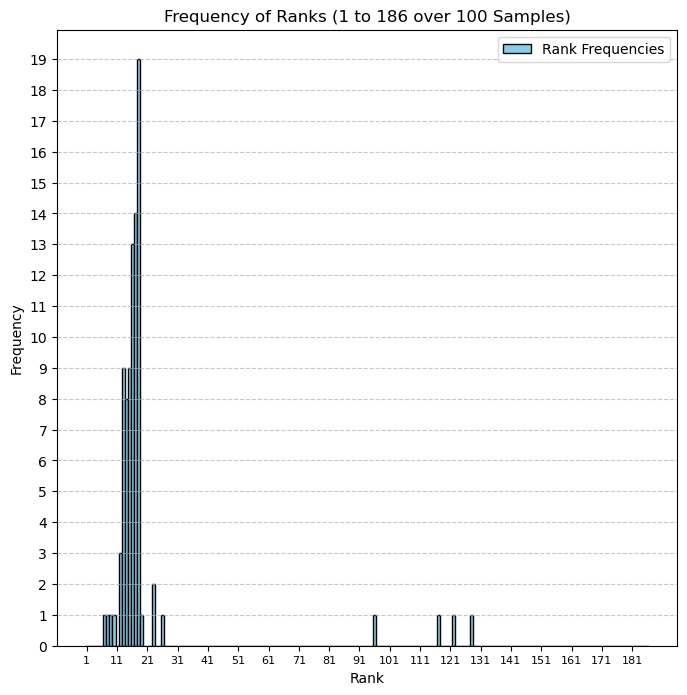}
    \label{fig:freq_ieee30}
  \end{subfigure}
  \caption{Evaluation and frequency analysis for IEEE-118 case study across 100 samples.}
  \label{fig:ieee188_analysis}
\end{figure*}

As illustrated in Fig.~\ref{fig:ieee6_analysis}, the generated contingencies of the IEEE-6 system are ranked below the threshold value. They are distributed evenly from the $1^{st}$ to the $3^{rd}$ rank, and a dozen of them fall to the higher rankings. To this end, it can be said that DDPM-CS is able to generate the worst contingencies for any base case of load demand in the $[0.5, 1.5]$ range of the common load profile based on what it has learned from the training dataset with not too many samples. 


Similarly, the IEEE-14 case study's simulation in Fig.~\ref{fig:ieee14_analysis} shows that the generated contingencies are ranked around the $1^{st}$ and $2^{nd}$ compared to the other ranks. The frequency of the $2^{nd}$ rank is the highest, followed by the $1^{st}$ ranks; meanwhile, the $6^{th}$ ranks are rare. The simulation results indicate that the Denoising Diffusion Probabilistic Models for Contingency Screening (DDPM-CS) performs better in the case of the IEEE-14 case study. It can be understood that the total number of contingencies is more than the IEEE-6 case studies, and the DDPM-CS remains effective in learning from the training dataset, though this case study's scale is larger than that of IEEE-6.

Regarding the IEEE-30 system, the simulations for the case studies are depicted in Fig.~\ref{fig:ieee30_analysis}. The simulation shows that the ranking of generated contingencies hovers around the $1^{th}$ and the $5^{th}$ ranks over the $38$ possible contingencies. Although the outcome is prone to less efficient compared to the previous ones due to the increasing complexity of the case study, the ranking of the generated contingencies is far below the threshold value, and close to the worst contingency. In particular, the number of contingencies that are ranked higher than the threshold is dominant compared to the smaller-scale case study and concentrated at the top of ranking. The largest frequency belongs to the $2^{nd}$ ranking with $41$ generated contingencies, followed by the $1^{st}$ ranking with $28$ contingencies; meanwhile, the remaining rankings account for only a negligible number of cases. 

Finally, scaling up the case study to the IEEE-118 system, more than four times larger than the IEEE-30 system, provides a meaningful test of how the proposed model DDPM-CS performs as system complexity increases substantially. The simulation results in Fig.~\ref{fig:ieee188_analysis} indicate that the generated contingencies are most frequently ranked between the $5^{th}$ and the $20^{th}$ positions among the $186$ possible contingencies. The frequency of the cohort $10^{th}-15^{th}$ rank is the dominant group, followed by the remaining, inconsiderable group. These observation suggests that the DDPM-CS exhibits reduced efficiency on the IEEE-118 system compared to the smaller test cases. Even so, the proposed model continues to demonstrate effective learning behavior and maintain reasonable performance despite the significantly larger scale and complexity relative to the previous ones.

Briefly, the performance of the proposed model DDPM-CS is summarized in Table~\ref {tab:sum_performance}. The 50\%-below ratio is the percentage of generated contingencies that are ranked below the threshold value, which is the median of the total number of possible contingencies per case study. The number of $N-1$ possible contingencies is the total number of contingencies that are able to be generated by the algorithm \ref{alg:1} for each case study. The results demonstrate that the DDPM-CS performs well in generating the worst contingencies for all case studies despite the increasing complexity via the shifting of the system scale. In other words, the learned data distribution is well-approximated to the true data distribution when it can capture the hidden relationship pattern between the worst contingencies and the system's power profiles. It is essential to include the dynamic nature of load demand in power systems for contingency anticipation in a manner that is less time-consuming and has a light computational burden through the quick inference of the proposed model. 

The resulting performance is a preliminary achievement for an application of diffusion theory in addressing complicated problems in power systems. There remain several issues to deal with in the proposed model in its application in the real world. The dominant one is that the convergence speed is no longer as fast as the scale of the case study increases. Because of the two-dimensional input structure stacked from the systems information, the model dimensionality of feature spaces rises as an exponential function of the system size. Additionally, the model training algorithm uses stochastic gradient descent, which leads to slow convergence, as observed in experiments. A robust high-performance infrastructure seems essential for scaling up larger case studies, and advanced techniques to accelerate the convergence speed of the training process are expected to be developed in future work.

\begin{table}[!htbp]
\caption{Summary of the performance of the proposed DDPM-CS model in generating contingencies, where lower rankings correspond to more detrimental contingencies.}
\centering
\setlength{\tabcolsep}{14pt} 
\renewcommand{\arraystretch}{1.5} 
\begin{tabular}{c c c}
\hline
\textbf{Case Study} & \textbf{50\%-below ratio} & \textbf{Number of N-1} \\ 
                    &                           & \textbf{possible contingencies}  \\ \hline 
\textbf{IEEE-6}     & 100\%                     & 11                          \\ \hline
\textbf{IEEE-14}    & 100\%                     & 18                          \\ \hline
\textbf{IEEE-30}    & 99\%                      & 38                          \\ \hline
\textbf{IEEE-118}   & 83\%                      & 186                         \\ \hline
\end{tabular}
\label{tab:sum_performance}
\end{table}



Based on the achieved results, the proposed Denoising Diffusion Probabilistic Models for Contingency Screening (DDPM-CS), grounded in diffusion theory, demonstrates reliability and robustness as a novel approach for alleviating the computational burden associated with identifying the most critical scenarios in power systems. Notably, the  DDPM-CS is a physics-aware generative model that leverages physical information from a specific power profile to generate a line outage, which is likely to pose significant risks to the power system's stability. The physical information from a current operating point is taken as a prompt to guide the model in predicting worst contingencies. Thereby, it provides system operators with actionable insights without the need for costly and time-consuming recalculation of all possible scenarios from the current operating point as required by traditional numerical methods. The physics-informed learning paradigm allows the model to address the stringent constraints of physical operational conditions and accurately capture the underlying relationship between the most detrimental contingencies and the current operating system points when the system structure varies. 

\section{Conclusion}
\label{conclusion}

This paper proposes a novel and unprecedented approach to deal with intensive efforts that must be consumed when working with contingency analysis by traditional methods. Instead of manipulating traditional numerical methods that need the outcome of power flows in systems to rank contingencies, the worst contingencies are generated by a generative machine learning model, named as the Denoising Diffusion Probabilistic Model for Contingency Screening (DDPM-CS). The diffusion mechanism is leveraged and adapted in DDPM-CS to align with the intrinsic complexity of the power system. The proposed model is physically informed by some worst contingencies determined beforehand to learn distinctive data patterns before generating a novel sample that is not different from the expected outcome of contingency screening.


This work may be considered as a preliminary-pioneered study of generative AI's applications in power system operation. It suggests a unique solution to adapt the famous generative AI model for image processing (i.e., stable diffusion) to the power system domain and proves its judiciousness by the simulation results. Although it is conducted on the four typical case studies due to the hardware limitation of experiments, its consistent simulation results are undeniable evidence of the proposed model's correctness, effectiveness, and scalability. Looking ahead, the DDPM-CS holds a promise as an alternative solution to conventional approaches, with potential for further.


%




\ifCLASSOPTIONcaptionsoff
  \newpage
\fi



%
\bibliographystyle{IEEEtran}
\bibliography{IEEEabrv,bibtex/bib/IEEE}
\end{document}